\documentclass[pra,superscriptaddress,amsmath,amssymb,twocolumn]{revtex4-2}

\usepackage{graphicx}  
\usepackage{color}
\usepackage{dcolumn}
\usepackage{bm}

\usepackage[colorlinks,bookmarks=true,citecolor=blue,linkcolor=blue,urlcolor=blue]{hyperref}

\begin{document}

\preprint{AIP/123-QED}

\title{
    Dynamical localization of interacting ultracold atoms in one-dimensional quasi-periodic potentials
}

\author{Áttis V. M. Marino}
\affiliation{Instituto de Física de São Carlos, Universidade de São Paulo, CP 369, 13560-970 São Carlos, SP, Brazil.}
\author{M. A. Caracanhas}
\affiliation{Instituto de Física de São Carlos, Universidade de São Paulo, CP 369, 13560-970 São Carlos, SP, Brazil.}
\author{V. S. Bagnato}
\affiliation{Instituto de Física de São Carlos, Universidade de São Paulo, CP 369, 13560-970 São Carlos, SP, Brazil.}
\affiliation{Department of Biomedical Engineering, Texas A\&M University, College Station, Texas 77843, USA}
\author{B. Chakrabarti}
\email{barnali.physics@presiuniv.ac.in}
\affiliation{Instituto de Física de São Carlos, Universidade de São Paulo, CP 369, 13560-970 São Carlos, SP, Brazil.}
\affiliation{Department of Physics, Presidency University, 86/1   College Street, Kolkata 700073, India.}
\date{\today} 
\begin{abstract}
We present numerically exact non-equilibrium dynamics of a one-dimensional Bose gas in quasi-periodic lattice that plays an intermediate role between the long-ranged order and truly disordered systems exhibiting unusual correlated phases. Precision control over lattice depth, interaction strength and filling factor enables the exploration of various correlated phases in a finite periodic lattice. We investigate the system’s dynamics when the secondary incommensurate lattice is abruptly switched on. To solve the many-body Schr\"odinger equation, we employ the multiconfigurational time-dependent Hartree method for bosons (MCTDHB). The many-body dynamics are analyzed through distinct measures of the Glauber correlation functions and dynamical fragmentation. Our study reveals four distinct scenarios of localization process in the non-equilibrium dynamics. Weakly interacting non-fragmented superfluid of incommensurate filling in the primary lattice exhibits collapse-revival dynamics of localization. In contrast, a fragmented superfluid with commensurate filling exhibits dynamical Mott localization. A strongly correlated, fully fragmented Mott state shows a subtle competition with localization introduced by the secondary lattice that merely melts the Mott correlations. Interestingly, in the fermionized Mott regime, where the density in each well is fragmented, the intra-dimer correlations exhibit unexpected robustness. These findings provide new insights into many-body correlation dynamics and novel localization mechanisms in quasi-periodic lattices, paving the way for engineering exotic quantum behaviors in ultracold atomic systems.
\end{abstract}


\makeatother
\maketitle

\section{Introduction} \label{sec:intro}

In the field of ultracold gases, the experimental advances in controlling optical lattices have enabled the investigation of one of the most remarkable paradigmatic phase transitions ---from a coherent superfluid (SF) phase to a localized Mott-insulator (MI) phase~\cite{nature.415,Spielman:2007,Esslinger:2004}. However, quantum gas physics took a new direction following the realization of Anderson localization in ultracold matter waves by Billy {\it et. al}~\cite{Billy:2008}, subsequently confirmed by other groups~\cite{Roati:2008, Fallani:2005, Fallani:2007, Modugno:2010}. Since then, the study of ultracold atoms in quasi-crystalline optical lattice has emerged as a highly promising platform for exploring exotic quantum phases. Experimentally, quasi-crystalline optical lattices can be engineered either by a bichromatic potential~\cite{Lewenstein:2005,Kuhn:2005} or by random potential~\cite{Fallani:2005, Fallani:2007}. A bichromatic optical lattice is formed by a primary optical lattice of higher intensity, superimposed with a weaker secondary lattice of incommensurate period. The secondary lattice thus effectively introduces disorder in the on-site energies of the primary lattice. Quasi-periodic potential plays a key role at the interface of truly long-range order and genuine disorder. Various exotic phenomena, such as Bose glass (BG) phase~\cite{Giamarchi:1988, Fisher:1989, Palencia:2019, Pasienski_2010}, many-body localization~\cite{Bloch:2019, Vadim}, fractal Mott lobes~\cite{Palencia:2019, Palencia:2020, Yao_2024}, topological phase transition ~\cite{Sudeshna} and Anderson localization~\cite{Lugan:2007, Lugan:2011, Lugan1:2007} are observed in quasi-periodic potential. Theoretically, the phases in the bichromatic optical lattice are usually studied using the Bose-Hubbard model~\cite{Jaksch:1998, Pinaki:2007, Noor:2023}, numerical solutions of the Gross-Pitaevskii (GP) equation~\cite{Adhikari:2009, Damski:2003, Lewenstein:2005, Kuhn:2005}, Quantum Monte-Carlo simulations~\cite{Pinaki:NJP, Richard:1991, Moore:2018}, and the Density Matrix Renormalization Group (DMRG) method.~\cite{Zwerger:1999}. Phase diagrams in quasi-periodic lattices have been investigated theoretically in both one- and two-dimensional systems~\cite{Zhu, Roux, Yao:2021, Dalfovo:2009, Dalfovo:2011, Dalfovo:2019}.

Loading ultracold atoms into 1D quasi-periodic lattices opens the possibility of high-precision quantum simulations of disordered interacting systems. These 1D setups provide a versatile framework to investigate the delicate balance between interaction-induced correlations and disorder-induced localization. In the case of commensurate filling in periodic lattice, the system realizes an incompressible Mott-insulator (MI) phase, while incommensurate filling typically supports an extended superfluid (SF) phase~\cite{rhombik_pra2018, rhombik_pra2025}. By tuning lattice depth or interaction strength, a MI–SF transition can be observed. However, in the presence of secondary lattice, this transition is interrupted by a compressible yet non-coherent Bose glass (BG) phase~\cite{Palencia:2020,Chiara,Chiara1,Tanzi_2013,Meldgin:2016,Fangzhao:2021}. 

The competition between correlation and localization becomes even more intriguing in strongly interacting bosons in 1D quasi-periodic lattices. Sophisticated experimental techniques—such as optical pumping~\cite{op-pump1, op-pump2}, geometric lattice engineering~\cite{gc}, evaporative cooling~\cite{ec}, and optical tweezers~\cite{ot} have made it possible to isolate and control highly correlated few-body systems. In these finite-sized platforms, quantum fluctuations are significantly enhanced due to reduced particle numbers and pronounced finite-size effects, making them ideal for probing microscopic correlations and emergent quantum phases. Recent studies highlight the emergence of novel behavior, where Mott-type correlations compete directly with disorder-induced correlations—ushering in a new regime of physics beyond the traditional mean-field and Bose-Hubbard descriptions~\cite{Yao_2024, BarnaliPRB, Molignini:2025}.

While considerable progress has been made in clean finite sized lattice systems in the strongly interacting limit, the role of secondary lattice remains less understood, partly due to the lack of theoretical tools capable of capturing strongly correlated dynamical features~\cite{Mi2017}. The key question is the dynamical stability or transition when the secondary detuning lattice is suddenly switched on. In this work, we simulate a realistic experimental protocol in which ultracold atoms are first prepared in a primary optical lattice and then suddenly subjected to a secondary incommensurate lattice, allowing the system to evolve in time into a quasi-periodic potential in 1D. We employ the multiconfigurational time-dependent Hartree method for bosons (MCTDHB)~\cite{Streltsov:2006, Streltsov:2007, Alon:2007, Alon:2008, Lode:2016,Lode:2020}, which goes beyond the mean-field and Bose-Hubbard (BH) frameworks~\cite{Jaksch:1998, RevModPhys.80.885, Fisher:1989, Esslinger:2004}, enabling a fully \emph{ab initio} treatment of the many-body dynamics. While the BH model describes the superfluid (SF) to Mott insulator (MI) transition in terms of the ratio between interaction strength and tunneling ($\frac{U}{J}$), its applicability is limited to site-localized Wannier states~\cite{Astrakharchik:2016, Sascha:2010} and weak correlations. In contrast, MCTDHB captures the SF phase via macroscopic occupation of a single orbital with long-range coherence, and the MI phase through orbital fragmentation and suppressed inter-well correlations. By preparing four distinct ground-state phases in the clean lattice and tracking their evolution after the quench in the secondary lattice, we probe the transition from disorder-free localization/delocalization to eventual localization in the quasi-crystalline potential. The dynamics of one- and two-body correlations reveal the interplay between interaction-induced and disorder-induced localization mechanisms in regimes previously inaccessible to standard approaches.

In a finite optical lattice, by independently controlling the interaction strength, lattice depth, and filling factor, it is possible to prepare initial configurations that differ not in their one-body densities, but rather in their initial correlations. The concept of orbital fragmentation extends the knowledge of \emph{non-fragmented superfluid} states of weakly interacting atoms with incommensurate filling to \emph{fragmented and strongly correlated superfluid} with commensurate filling. Whereas strongly interacting atoms in deep lattice can either be \emph{fully-fragmented Mott} states with unit filling or \emph{fermionized Mott} states with double filling, demonstrating physics beyond the BH model in the periodic lattice. These four distinct initial setups strongly differ in their initial correlations and exhibit stringent difference in dynamical evolution in the secondary lattice. When the non-fragmented SF exhibits a \emph{collapse-revival scenario in the dynamical localization}, the fragmented SF displays \emph{dynamical Mott localization}. Fully-fragmented Mott exhibits a complex interplay between Mott correlations and disorder-induced localization resulting in the \emph{melting of Mott correlations}.
Whereas the fermionized Mott remains \emph{robust} even in the presence of strong disorder.

We study the dynamical evolution and the subsequent stability of each phase by calculating observables such as one- and two-body Glauber correlation functions. The intricate interplay between correlations in the primary periodic lattice and the disorder-induced correlations in the quasi-periodic lattice is further studied by orbital fragmentation and the order parameter, unveiling novel dynamical mechanism.

The structure of the article is as follows. In Sec.~\ref{sec:protocol}, we present the many-body Hamiltonian and describe the quench protocol. Sec.~\ref{sec:methods} outlines the methodology and the physical observables of interests. In Sec.~\ref{sec:initial}, we describe the initial setups and analyze the prequench one-body densities. Sec.~\ref{sec:quench} discusses the disorder-induced dynamics in the quasi-periodic potential and evaluates the Glauber correlation functions for the four distinct initial configurations. Finally, Sec.~\ref{sec:conclusion} provides a summary and conclusions.

\section{System and Protocol} \label{sec:protocol}

The dynamics of $N$ interacting bosons of mass $m$ in a one-dimensional optical lattice is governed by the time-dependent Schrödinger equation:
\begin{equation}
    \hat{H} \psi = i \hbar \frac{\partial \psi}{\partial t}.
\end{equation}

The total $N$-body Hamiltonian has the form:
\begin{equation} 
    \hat{H}(x_1,x_2, \dots x_N)= \sum_{i=1}^{N} \hat{h}(x_i) + \sum_{i<j=1}^{N}\hat{W}(x_i - x_j).
\label{propagation_eq}
\end{equation}

Its one-body part is $\hat{h}(x) = \hat{T}(x) + \hat{V}(x)$, where $\hat{T}(x) = -(\frac{\hbar^2}{2m}) \frac{\partial^2}{\partial x^2}$ is the kinetic energy operator. $V(x)$ is the quasi-periodic lattice, which is realized by superposing a detuning laser of amplitude $V_d$ and wave vector $k_d$ with a primary laser of amplitude $V_p$ and wave vector $k_p$. 
\begin{equation}
 V(x) = V_p \sin^2(k_p x) + V_d \sin^2 (k_d x).
\end{equation}

$V_p$ and $V_d$ are measured in units of recoil energies: $E_{r_p} = \frac{\hbar^2 k_p^2}{2m}$ and $E_{r_d} = \frac{\hbar^2 k_d^2}{2m}$. The external potential generates a regular but non-repeating structure that implements correlated disorder on top of the periodic primary lattice when the ratio of the two wave vectors $\frac{k_d}{k_p}$ is chosen to be incommensurate. The actual experiment operates with wavelengths $\lambda_p = \frac{2\pi}{k_p} = 1032$~nm and $\lambda_d = \frac{2\pi}{k_d} = 862$~nm, leading to an incommensurate ratio $\frac{k_d}{k_p} \simeq 1.1972157773\dots$~\cite{Roati:2008, Adhikari:2009}. We choose $k_p$ and $k_d$ following~\cite{Roati:2008, Adhikari:2009}, which takes the inverse of the golden ratio $\Phi = \frac{(\sqrt{5}-1)}{2}$ as the target incommensuration parameter. However, due to finite numerical precision, the incommensurate frequency ratio used to generate the quasi-periodic potential will necessarily be approximated by a rational number. We have verified that this approximation preserves the quasi-periodic features within the limits of the numerical grid. Thus, in the entire numerical simulation, we keep the wave vectors as $k_p = 1.0$ and $k_d = 1.197215$.

The bosons interact through two-body short-range interaction $W(x_i - x_j) = g_0 \delta(x_i - x_j)$ with interaction strength $g_0 > 0$ in the entire calculation. We remark that the Hamiltonian $\hat{H}$ can be written in dimensionless units obtained by dividing the dimensionful Hamiltonian by $\frac{\hbar^2}{m{\bar{L}}^2}$, with $\bar{L}$ an arbitrary length scale, which, for the purposes of our calculations, is set to the period of the optical lattice. We probe various values of the potential depth $V_p \in [5 E_r, \, 15 E_r]$  and $V_d \in [0.2 E_r, \, 5.0 E_r]$, where $E_r = \frac{\hbar^{2}k_p^{2}}{2m}$ is the recoil energy of the primary lattice. To probe the full pathway of dynamical localization, we consider particles with interaction strengths ranging over four orders of magnitude, from very weak to strong interactions in the range of $g_0 \in [0.001 E_r, \, 10 E_r]$. The geometry is restricted to $S$ sites in the optical lattice, with the spatial extent defined between two maxima of the sinusoidal potential. Depending on the number of particles
considered, we restrict the system geometry to accommodate
$S = 7$ or $S = 9$ or $S=3$ sites in the primary optical lattice. To ensure a finite-sized primary lattice, hard-wall boundary conditions are applied. Possible setups to probe the physics of both incommensurate and commensurate filling include: incommensurate filling ($N=10$ bosons in $S=7$ sites); commensurate filling (filling factor one, $N=9$ bosons in $S=9$ sites); commensurate filling (filling factor two, $N=6$ bosons in $S=3$ sites). In the rest of discussion, we describe the strength of the secondary lattice as "disorder" for simplicity.

In the quench protocol, we first obtain the ground state for the Hamiltonian in the primary lattice ($V_d = 0 E_r$), then at time $t > 0$, we propagate the state by quenching it to the disordered lattice of various strength $V_d$.

\section{Methods} \label{sec:methods}

To simulate the dynamics of the full many-body interacting systems, we employ multi-configurational time-dependent Hartree method for indistinguishable particles (MCTDHB)~\cite{Streltsov:2006, Streltsov:2007, Alon:2007, Alon:2008, Lode:2016, Fasshauer:2016, Lode:2020}  coded in the MCTDH-X software~\cite{Lin:2020, MCTDHX}. MCTDH-X solves the many-body Schrödinger equation through a time-dependent variational optimization procedure. The many-body wave function is decomposed into an adaptive basis of $M$ time-dependent single-particle functions, referred to as orbitals. Both the expansion coefficients and the orbitals are optimized in time, either to obtain the ground state via imaginary time propagation or to compute the full time evolution through real-time propagation.

The wave function of the interacting $N$-boson is expanded over a set of time-dependent permanents:
\begin{equation}
    \left| \Psi(t) \right >= \sum_{\mathbf{n}} C_{\mathbf{n}}(t) \vert \mathbf{n}; t \rangle.
\label{many_body_wf}
\end{equation}

Consequently, the permanents are constructed over $M$ time-dependent single-particle wavefunctions, called orbitals, as:
\begin{equation}
    \vert \mathbf{n}; t \rangle = \prod_{k=1}^M \left\{ \frac{[\hat{b}_k^\dagger(t)]^{n_k}}{\sqrt{n_k !}} \right\} \vert 0 \rangle.
\label{many_body_wf_2}
\end{equation}

Here, $\mathbf{n} = (n_1, n_2, \dots, n_M)$ is the number of bosons in each orbital. This is constrained by $\sum_{k = 1}^M n_k = N$, with $N$ the total number of bosons. $|0\rangle$ is the vacuum state and $\hat{b}_k^\dagger(t)$ denotes the time-dependent operator that creates one boson in the $k$-th working orbital $\psi_k(x)$.

A formal variational treatment with the above ansatz leads to the MCTDHB equations of motion. Both the expansion coefficients $C_\mathbf{n}(t)$ and the working orbitals $\psi_i(x; t)$ that constitute the permanents are optimized variationally at every time step~\cite{TDVM81} to either relax the system to its ground state (imaginary time propagation) or to calculate the full dynamics of the many-body state (real time propagation). We require the stationarity of the action with respect to variations of both the time-dependent coefficients and the orbitals, which results in a coupled set of equations of motion for the coefficients and orbitals. These equations are solved simultaneously. It is important to note that both the single-particle functions and the coefficients are variationally optimized with respect to all parameters of the many-body Hamiltonian~\cite{TDVM81, variational1, variational3, variational4}.

For $M = 1$ (a single orbital), MCTDHB coincides with a mean-field Gross-Pitaevskii description. For $M \rightarrow \infty$, the wave function becomes exact as the set $\vert n_1, n_2, \dots, n_M \rangle$ spans the complete $N$-particle Hilbert space. However, for practical calculations, we restrict the number of orbitals, and thus, the accuracy of the algorithm depends on the number of orbitals used to achieve convergence in the relevant observables. Convergence is ensured when the population of the highest orbital is negligible and various dynamical measures, such as the density and correlation functions, become converged.

We calculate several observables from the many-body state $\left| \Psi(t) \right>$. To probe the spatial distribution of the bosons, we calculate the one-body density as:
\begin{equation}
    \rho(x; t) = \left< \Psi(t) \right| \hat{\Psi}^{\dagger}(x) \hat{\Psi}(x) \left| \Psi(t) \right>.
\end{equation}

To measure the degree of coherence and many-body correlation, we calculate one-body and two-body Glauber correlation functions, defined as:
\begin{align}
    g^{(1)}(x, x'; t) &= \frac{\rho^{(1)}(x, x'; t)}{N \sqrt{\rho(x; t) \rho(x'; t)}} \quad \\
    g^{(2)}(x, x'; t) &= \frac{\rho^{(2)}(x, x'; t)}{N^2 \rho(x; t) \rho(x'; t)}.
\end{align}
    
Whereas the reduced one-body and two-body densities are defined respectively as:
\begin{align}
    \rho^{(1)}(x, x'; t) &= \left< \Psi(t) \right| \hat{\Psi}^{\dagger}(x) \hat{\Psi}(x') \left| \Psi(t) \right>  \quad \\
    \rho^{(2)}(x, x'; t) &= \left< \Psi(t) \right| \hat{\Psi}^{\dagger}(x) \hat{\Psi}^{\dagger}(x') \hat{\Psi}(x') \hat{\Psi}(x) \left| \Psi(t) \right>.
\end{align}
 
From the reduced one-body density matrix, it is possible to obtain information regarding the orbital occupation throughout the time evolution, i.e. how much of the provided Hilbert space is dynamically occupied. This can be expressed via the natural orbitals $\phi_i^{(\mathrm{NO})}$ and the orbital occupations $\rho_i$, which are the eigenfunctions and eigenvalues of $\rho^{(1)}(x, x')$, i.e.
\begin{equation}
    \rho^{(1)}({x},{x}') = \sum_i \rho_i \phi^{(\mathrm{NO}),*}_i({x}')\phi^{(\mathrm{NO})}_i({x}).\label{eq:RDM1}
\end{equation}

When a single natural orbital is macroscopically occupied, the one-body density matrix $\rho^{(1)}$ has only one significant eigenvalue, and the many-body state is a non-fragmented condensate and can be accurately described by mean-field theory. When $\rho^{(1)}$ has $k$ macroscopically occupied eigenvalues ---the system is referred to as $k$-fold fragmented~\cite{onsager}.

We further define a mesoscopic ``order parameter'' using the eigenvalues of the reduced one-body density matrix as:
\begin{equation}
    \Delta(t) = \sum_{i} \left( \frac{n_i(t)}{N} \right)^2,
\end{equation}

where $n_i(t)$ is time-dependent natural occupation in $i^{\rm th}$ orbital. For the superfluid phase, $\Delta= 1$ as only one eigenvalue is non-zero. For the Mott phase, when the number of significantly contributing orbitals becomes equal to the number of sites $(S)$, $\Delta$ becomes $\frac{1}{S}$. Thus $\Delta= 1$ and $\Delta= \frac{1}{S}$ represent the two extreme limits corresponding to the superfluid and Mott-insulating phases, respectively. We find that the time-dependent order parameter smoothly interpolates between these two limits when disorder-induced Mott localization occurs. 

\begin{figure*}[tbh]
    \centering
    \includegraphics[width=0.95\textwidth]{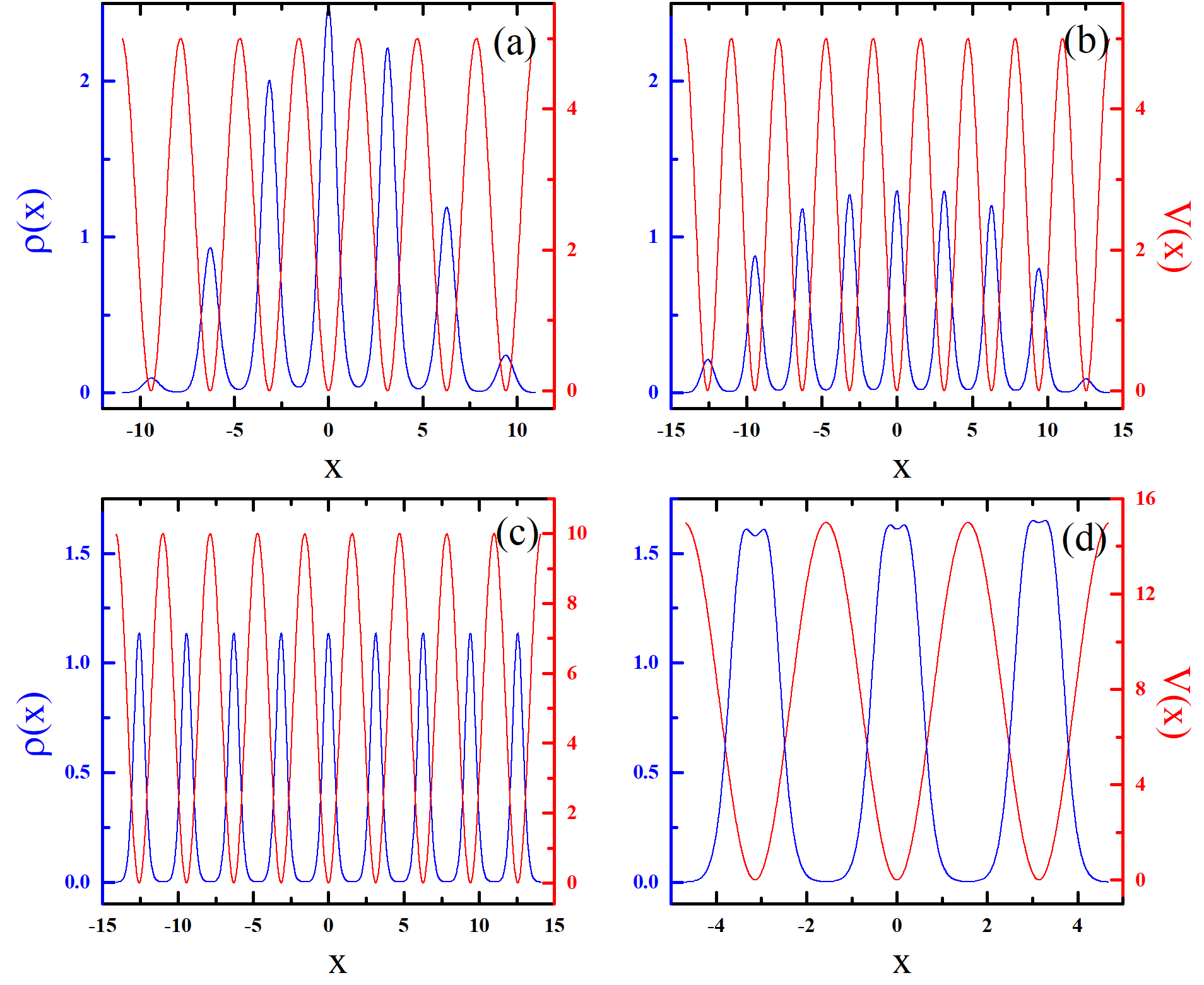}
    \caption{One-body density (solid blue line) $\rho(x)$ in the primary lattice for different initial setups. (a) incommensurate filling with $N = 10$ bosons in $S = 7$ lattice sites of lattice depth $V_p = 5 E_r$, interaction strength $g_0 = 0.001 E_r$, density is maximal in the central well, ground state is a non-fragmented superfluid phase. (b) commensurate filling (filling factor $\nu = 1$) with $N = 9$ bosons in $S = 9$ lattice sites of lattice depth $V_p = 5 E_r$, interaction strength $g_0 = 0.01 E_r$, ground state is a fragmented superfluid phase. (c) commensurate filling ($\nu = 1$) with $N = 9$ bosons in $S = 9$ lattice sites of lattice depth $V_p = 10 E_r$, interaction strength $g_0 = 0.5 E_r$, ground state is a fully fragmented Mott phase. (d) commensurate filling ($\nu = 2$) with $N = 6$ bosons in $S = 3$ lattice sites of lattice depth $V_p = 15 E_r$, interaction strength $g_0 = 10 E_r$, ground state is a fermionized Mott phase. Lattice potential is shown by red lines in all cases. See the text for details.}
    \label{fig:density}
\end{figure*}

\section{Density in the initial setups} \label{sec:initial}

Different aspects of localization of noninteracting and the interacting bosons in one-dimensional bichromatic lattice can be simulated. However, the response of interacting bosons in the disordered lattice is intimately related to the initial setups in the primary lattice sites. In the absence of disorder, depending on the strength of the periodic potential ($V_p$), interaction strength ($g_0$) and filling factor $\nu = \frac{N}{S}$, the ground state can belong to one of the four classes that are clearly distinguishable by their initial correlation. Fig.~\ref{fig:density}(a)-(d) summarizes the one-body density for four different initial setups. At sufficiently weak interactions, ($g_0 = 0.001 E_r)$, moderate lattice depth $V_p = 5 E_r$, and incommensurate filling with $N = 10$ bosons in $S = 7$ lattice sites, the ground state is a non-fragmented superfluid. In Fig.~\ref{fig:density}(a), we plot the corresponding one-body density, it is maximum in the central well and decreases as we go to the outer sites. Figs.~\ref{fig:density}(b)-(d) demonstrate the one-body density for commensurate filling factor. Fig.~\ref{fig:density}(b) exhibits the one-body density with filling factor $\nu = 1$, when $N = 9$ bosons are distributed in $S = 9$ lattice sites, we keep the lattice depth $V_p = 5 E_r$, but interaction strength is increased to $g_0 = 0.01 E_r$, the ground state is neither a superfluid nor a Mott insulator state, it is a fragmented superfluid (as determined by orbital fragmentation, which will be discussed later). To obtain the Mott state, we further increase the lattice depth to $V_p = 10 E_r$ and interaction strength to $g_0 = 0.5 E_r$. Fig.~\ref{fig:density}(c) exhibits Mott localization with a vanishing overlap of the density in the distinct well. We have checked that once localization happens, much stronger interaction no longer affects the density distribution (not shown here). Fig.~\ref{fig:density}(d) represents the one-body density for double filling case ($\nu = 2$). Keeping lattice depth and interaction strength as Fig.~\ref{fig:density}(c), we do not observe any onsite effect. To obtain fermionized Mott state and the formation of dimers, we increase the lattice depth to $V_p = 15 E_r$, whereas the interaction strength is also increased to $g_0 = 10 E_r$. However to keep the entire dynamics manageable with assured convergence, we reduce the number of sites to $S=3$ and number of bosons to $N=6$. The onset of fermionization is indicated by the appearance of a dip in each well, as two strongly interacting bosons residing in the same well attempt to avoid spatial overlap.

The four choices of prequench states in the primary lattice sites are significantly different in initial correlation and fragmentation (discussed later). Upon the sudden introduction of a secondary lattice potential, the system exhibits different responses, as the disorder-induced correlations compete in a complex manner with the initial correlations. We numerically investigate the full dynamics using one- and two-body correlation measures, as presented in the following four subsections.

\section{Quench dynamics in the quasi-periodic lattice} \label{sec:quench}

\subsection{Incommensurate filling and collapse-revival dynamics of localization}

\begin{figure}[tbh]
    \centering
    \includegraphics[width=1\columnwidth]{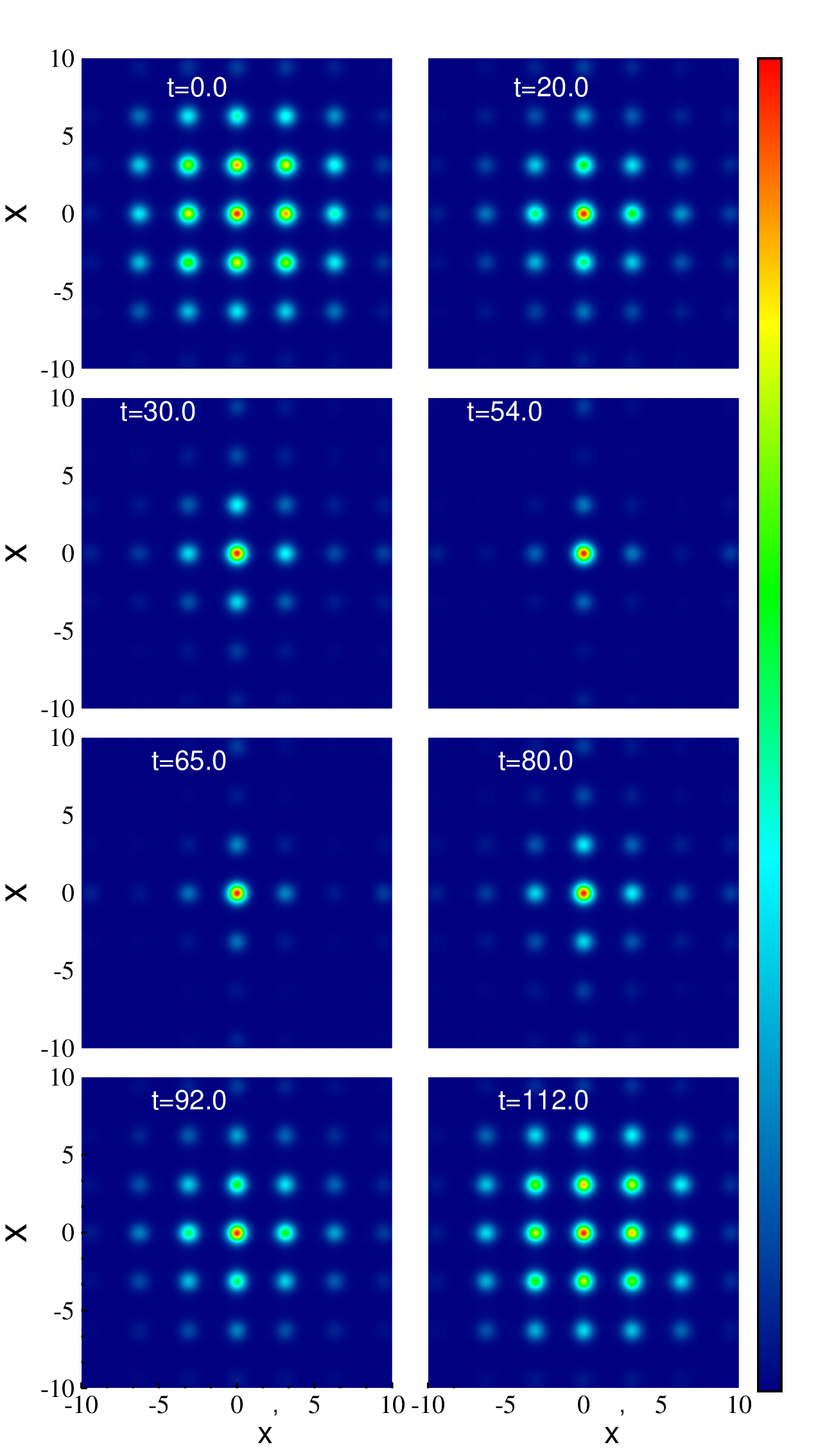}
    \caption{Dynamics of the reduced one-body density $\rho^{(1)}(x, x^{\prime})$ in the post quench state for $N = 10$ bosons in $S = 7$ lattice sites with repulsive interaction $g_0 = 0.001 E_r$, primary lattice depth $V_p = 5 E_r$ and orbital $M = 7$. In the quench dynamics, the superfluid phase in the clean lattice is suddenly quenched with disorder strength $V_d = 0.2 E_r$. One complete cycle of superfluid collapse to localization and subsequent revival is presented. See the text for details.}
    \label{fig:correlation-incomm}
\end{figure}


We begin by detailing the dynamical evolution of the non-fragmented SF in the quasi-periodic potential. We prepare the ground state of $N = 10$ weakly interacting bosons with $g_0 = 0.001 E_r$, in the $S = 7$ lattice sites of the primary lattice of depth $V_p = 5 E_r$ (Fig.~\ref{fig:density}(a)). In the sudden quench protocol, we turn on $V_d$. In Fig.~\ref{fig:correlation-incomm}, we plot the reduced one body density $\rho^{(1)}(x, x')$ for the weak disorder strength $V_d = 0.2 E_r$ at specific times, $t$ = 0, 20, 30, 54, 65, 80, 92 and 112 covering a complete cycle of collapse-revival (\emph{superfluid-localization-superfluid}). At time $t = 0$, long-range correlation is manifested in the nonvanishing off-diagonal behavior of $\rho^{(1)}(x, x')$. Although true off-diagonal long-range order is absent in such a finite ensemble, the off-diagonal elements of the one-body density matrix $\rho^{(1)}(x, x')$ still reflect correlations across the lattice, indicating that the initial state is a SF state. We also observe the off-diagonal correlation is gradually depleted with increasing distance from the central lattice. Introducing a small disorder, the off-diagonal correlations are progressively destroyed, ultimately leading to localization in the central and adjacent lattice sites around $t\approx 54$. As the atoms are weakly interacting, localization does not occur exclusively at the central site; the one-body density slightly spreads into the neighboring lattice sites. This is manifested by four faint spots surrounding the central bright spot. In the long-time dynamics, the off-diagonal coherence gradually starts to build up and we observe complete revival of long-range correlation across the lattice, leading to the superfluid phase at $t \approx 112$. The entire cycle is termed as \emph{collapse-revival dynamics of localization}. The corresponding two-body reduced density matrix $\rho^{(2)}(x, x')$ reflects the same physics (not shown here). Results obtained with a higher disorder strength, $V_d = 0.5 E_r$, also exhibit collapse–revival dynamics in the short time, however oscillation becomes dissipative in the longer timescale (not shown here).

\begin{figure}[tbh]
    \centering
    \includegraphics[width=1.0\columnwidth]{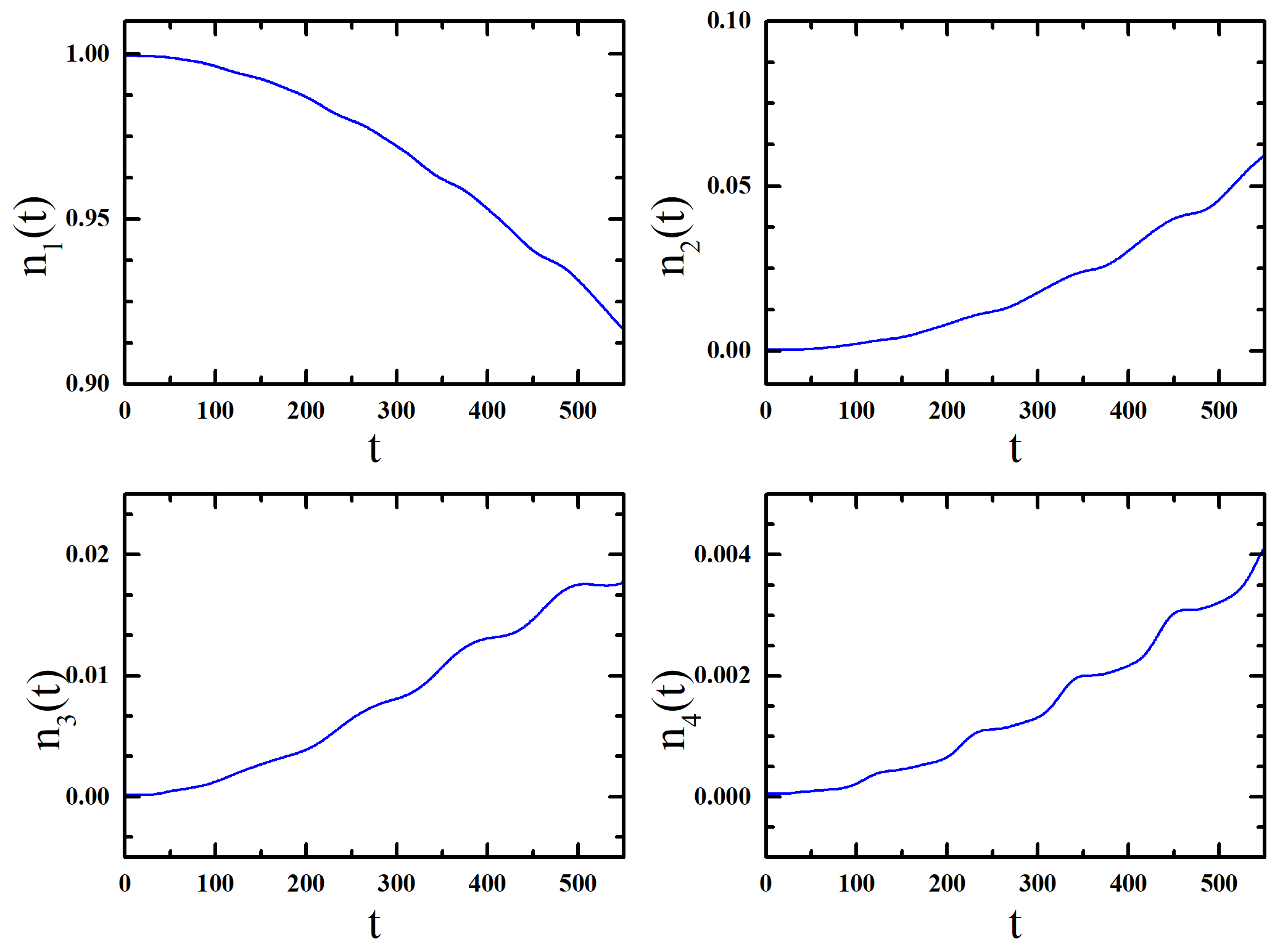}
    \caption{Dynamics of orbital fragmentation for the same quench protocol as in Fig.~\ref{fig:correlation-incomm}. The computation employs $M = 7$ orbitals, and the occupations of the lowest four natural orbitals with significant populations are presented. See the text for details.}
    \label{fig:fragmentation-incomm}
\end{figure}

Fragmentation is the hallmark of MCTDHB when more than one single particle state becomes significantly occupied. The dynamics of occupation across different orbitals provides a measure of dynamical fragmentation during the non-equilibrium evolution following a quantum quench. Fig.~\ref{fig:fragmentation-incomm} depicts the natural occupation as a function of time when the prequench non-fragmented superfluid state is suddenly quenched by the introduction of a secondary lattice of strength, $V_d = 0.2 E_r$. The computation is performed using $M = 7$ orbitals, and we plot the occupations of the lowest four orbitals during the long-time dynamics ($t=500$), since the contributions from the remaining three orbitals are negligible. Fig.~\ref{fig:fragmentation-incomm}(a) shows the occupation of the lowest and most significantly populated orbital. At $t = 0$, this orbital is fully occupied, while the other three orbitals (Figs.~\ref{fig:fragmentation-incomm}(b)-(d)) have zero occupation, indicating that the initial state is a non-fragmented superfluid describable by mean-field theory (i.e. $\vert N = 10, 0, 0, 0, 0, 0, 0, \rangle$). Over time, the occupation in the lowest orbital decreases, whereas the initially unpopulated three orbitals gradually become populated. Thus, the post-quench state exhibits dynamical fragmentation, although the degree of fragmentation is not strong: even at the final time point $(t = 500)$, the lowest orbital retains 94\% occupation, with the remaining 6\% distributed among the other orbitals. Moreover, the occupations of the four lowest orbitals exhibit oscillatory behavior rather than smooth dynamics. This wavy pattern is related with the time scale of the \emph{superfluid collapse–localization–superfluid revival} cycle (Fig.~\ref{fig:correlation-incomm}).

\subsection{Commensurate filling factor $\nu = 1$, dynamical Mott localization for fragmented superfluid}

For the commensurate filling factor, $\nu = 1$, with precision control over the lattice depth and interaction strength, the initial state can fall into one of three distinct regimes: i) For very weak interactions and moderate lattice depth, the ground state corresponds to a non-fragmented superfluid, exhibiting both inter- and intra-well coherence. ii) Increasing the interaction strength within the same lattice leads to a fragmented superfluid, in which several natural orbitals contribute significantly to the state. iii) For very strong interactions and deeper lattices, the system reaches a fragmented Mott-insulating state.

We first investigate the system using the same parameters as before (Section. A): primary lattice depth $V_p = 5 E_r$, weak interaction strength $g_0 = 0.001 E_r$, but now with commensurate filling: $N = 7$ bosons in $S = 7$ sites-the prequench state corresponds to a non-fragmented superfluid. The post-quench dynamics in the disordered lattice exhibit similar features to those observed in the incommensurate filling case discussed in the previous section. Repeating the calculation for larger lattice system size, $N=9$ bosons in $S=9$ lattice sites does not reveal any new physics.

\begin{figure}[tbh]
    \centering
    \includegraphics[width=1.0\columnwidth]{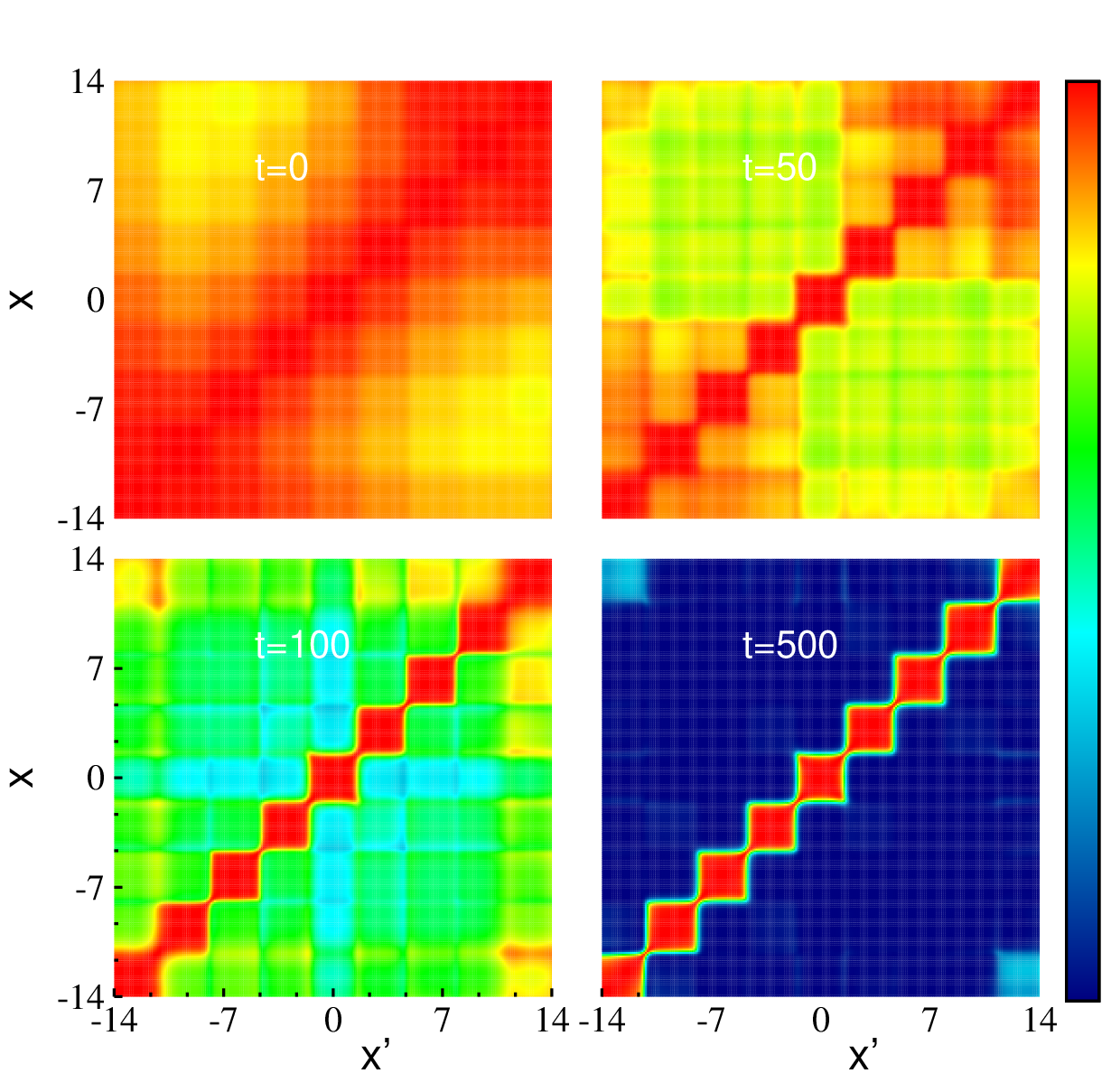}
    \caption{Dynamics of Glauber one-body correlation function $g^{(1)}(x, x')$ following a quench for $N = 9$ bosons in $S = 9$ lattice sites. The simulation is performed using $M = 10$ orbitals. Other parameters are: $V_p = 5 E_r$, $g_0 = 0.01 E_r$, $V_d = 0.5 E_r$. At $t = 0$, the prequench state exhibits complete first-order coherence across the lattice, characteristic of the superfluid phase. During the evolution, $g^{(1)}(x, x')$ develops nine bright lobes along the diagonal, with coherence maintained within each site and vanishing off-diagonal coherence ---signifying the onset of dynamical Mott localization. See the text for details.}
    \label{fig:corr-1B-SF}
\end{figure}

\begin{figure}[tbh]
    \centering
    \includegraphics[width=1.0\columnwidth]{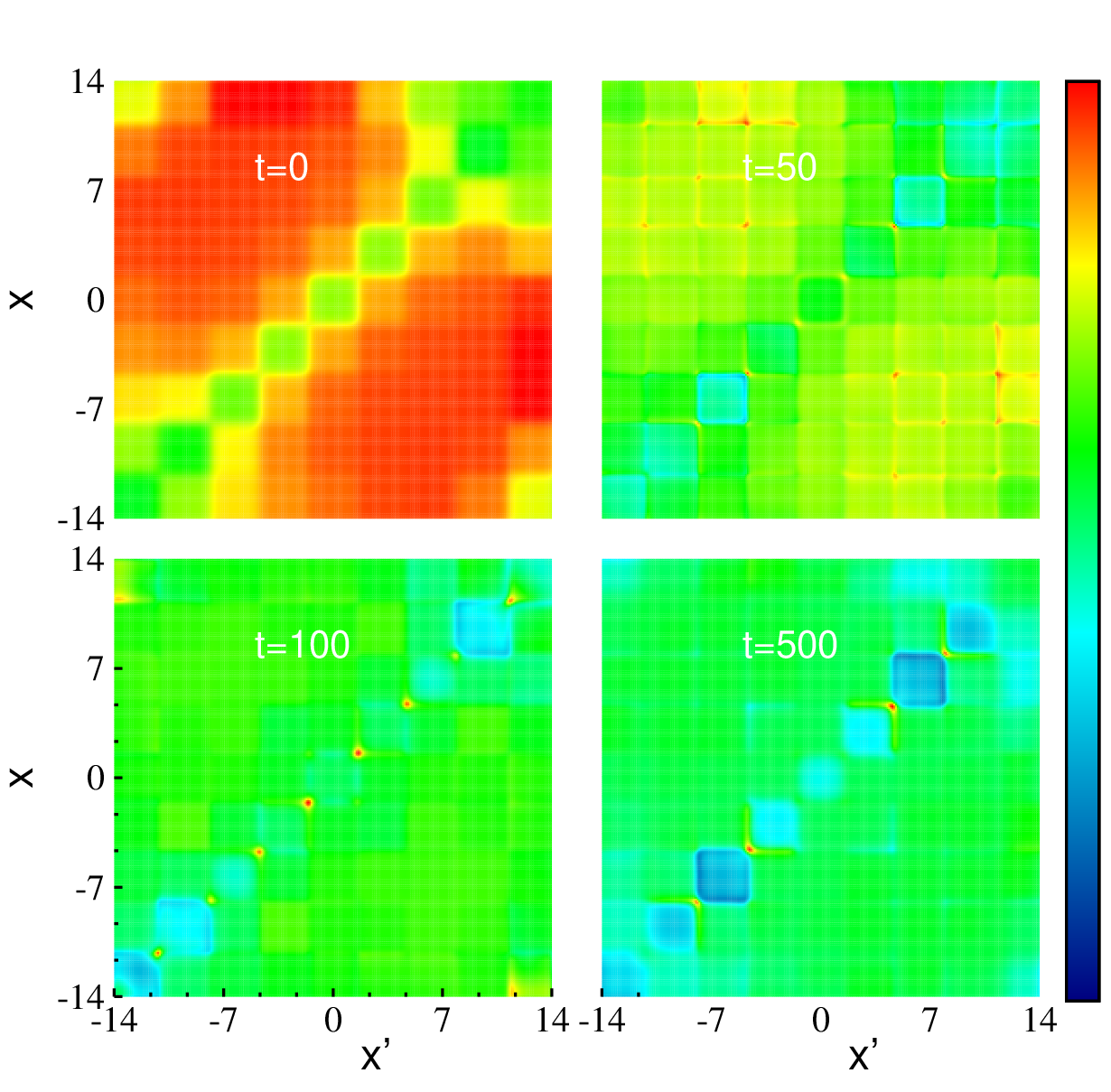}
    \caption{Dynamics of Glauber two-body correlation function $g^{(2)}(x, x')$ following a quench for the same parameters as in Fig.~\ref{fig:corr-1B-SF}. At $t = 0$, the prequench state exhibits complete second-order coherence between distinct wells, characteristic of the superfluid phase. At $t=500$, dynamical Mott localization, $g^{(2)}(x, x')$ reveals nine pronounced correlation holes—a strong suppression of two-body correlations along the diagonal—while off-diagonal two-body coherence remains. See the text for details.}
    \label{fig:corr-2B-SF}
\end{figure}

Next, keeping the lattice parameters same, but increasing interaction strength to $g_0 = 0.01 E_r$, the prequench state becomes fragmented SF (Fig.~\ref{fig:density}(b)), exhibiting correlation across the lattice but several orbitals contribute to the ground state (discussed later). The system’s response to a quench with disorder strength $V_d = 0.5 E_r$ is presented in Fig.~\ref{fig:corr-1B-SF} and Fig.~\ref{fig:corr-2B-SF}. In Fig.~\ref{fig:corr-1B-SF}, we plot the dynamics of the normalized first-order Glauber correlation function $|g^{(1)}(x, x')|^2$. At $t = 0.0$, we observe that coherence is preserved both within and between lattice sites. Complete first-order coherence is observed within each well, with $|g^{(1)}|$ $\simeq$ 1 for all $x \simeq x'$. Significant interwell coherence is also evident from the off-diagonal elements, where $x \neq x'$. Following the sudden introduction of the secondary lattice the off-diagonal correlations gradually diminish, while the diagonal correlations become increasingly dominant. At time $t = 500$, the diagonal of the correlation matrix shows nine completely separated coherent regions where $|g^{(1)}|^2 \simeq 1$, indicating that coherence is maintained within each individual well, but lost between distinct wells. This configuration corresponds to a nine-fold fragmented, fully localized Mott phase. The associated evolution is referred to as \emph{dynamical Mott localization}.

In Fig.~\ref{fig:corr-2B-SF}, we analyze the second-order coherence function $g^{(2)}(x', x, x', x) \equiv g^{(2)}(x, x')$ for the same selected time points as in Fig.~\ref{fig:corr-1B-SF}. At $t = 0$, second-order coherence between different wells is preserved, with $g^{(2)}(x, x') \simeq 1$ for the off-diagonal elements. The diagonal elements are slightly suppressed due to the anti-bunching effect, which arises from the repulsive interaction between particles. Throughout the dynamics, we observe the process of dynamical Mott localization, characterized by the gradual suppression of the diagonal elements of the normalized two-body correlation function. As the probability of double occupancy in a single well decreases, a correlation hole begins to emerge along the diagonal, while inter-well second-order coherence is maintained. By $t = 500$, nine distinct and diminished lobes appear, indicating the onset of \emph{dynamical Mott localization}. 

\begin{figure}[tbh]
    \centering
    \includegraphics[width=1.0\columnwidth]{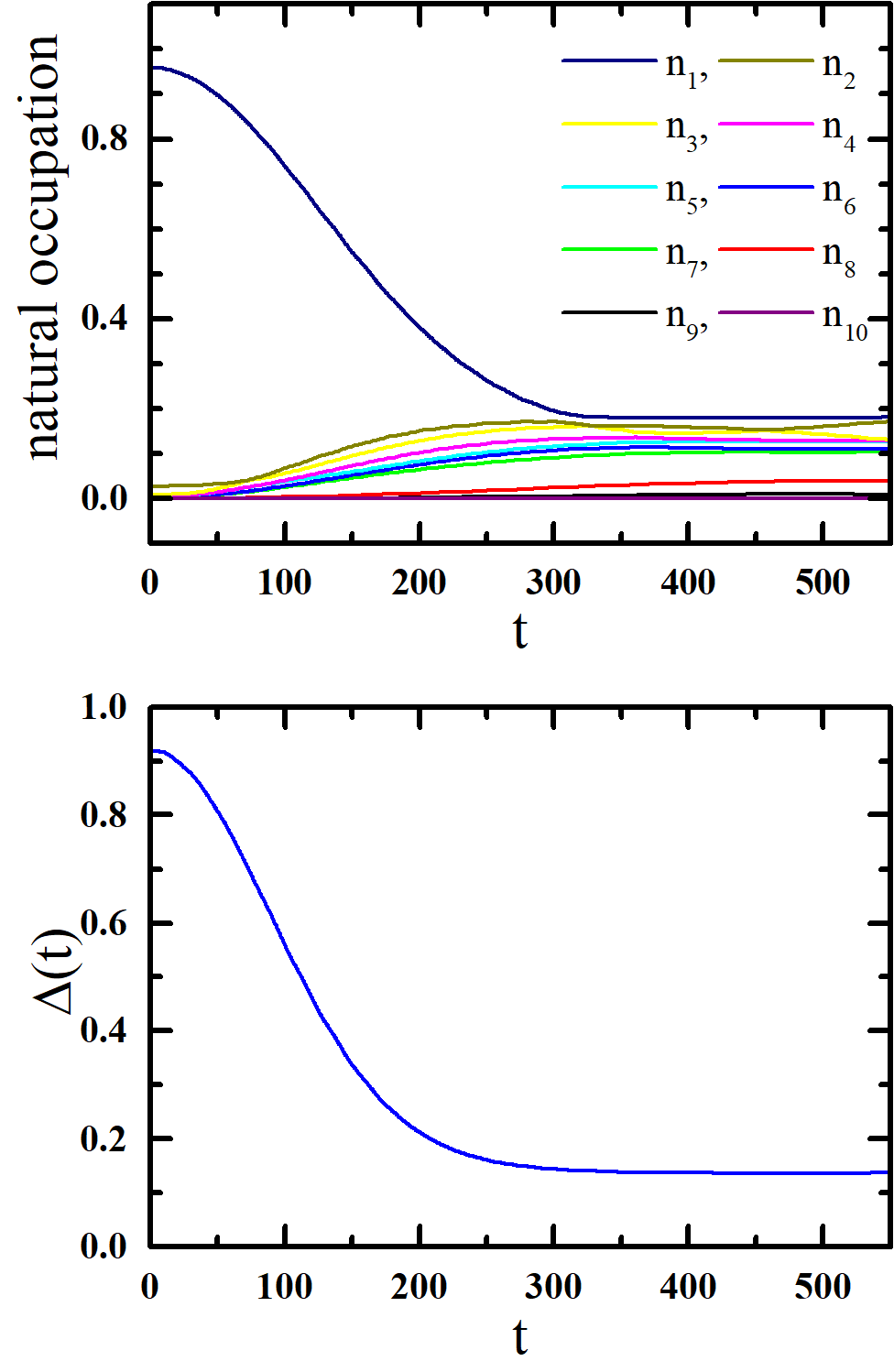}
    \caption{Top panel: Dynamics of orbital fragmentation following a quench from a fragmented superfluid, quench protocol is same as of Fig.~\ref{fig:corr-1B-SF}. At the point of dynamical Mott localization, maximal dynamical fragmentation is observed, with multiple orbitals contributing \emph{nearly equally}, but fully fragmented Mott is not achieved. Bottom panel: Time evolution of the order parameter $\Delta$, which decreases smoothly from $0.9$ to $0.14$, indicating a continuous transition from a fragmented superfluid to a Mott-localized state. See the text for details.}
    \label{fig:fragmentation-SF}
\end{figure}

In the top panel of Fig.~\ref{fig:fragmentation-SF}, we present the dynamical fragmentation throughout the entire evolution, computation is done using $M = 10$ orbitals. The occupations of all natural orbitals are plotted as functions of time. Initially, at $t = 0.0$, the occupation of the first and most significantly populated orbital is approximately 95\%, which is a macroscopic occupation. However, the superfluid cannot be considered a fully coherent mean-field state, as several other natural orbitals also contribute ---characterizing it as a fragmented superfluid. As time progresses, the population in the first orbital decreases, while the occupations of other orbitals increase, signaling the onset of dynamical fragmentation. At time $t = 350$, seven orbitals contribute with almost equal weights, with populations ranging from $0.11$ to $0.18$. This corresponds to a maximally dynamically fragmented Mott state. However, disorder alone does not induce a fully fragmented Mott state, which would require equal occupation of all nine orbitals. In the bottom panel, we show the time evolution of the order parameter throughout the simulation. At $t = 0$, $\Delta = 0.9$, indicating that the prequench state is a fragmented superfluid. In the dynamical evolution, $\Delta$ gradually decreases reaching a value of $0.14$ at $t = 350$, consistent with the emergence of dynamical Mott localization. However, since the disorder induces only a maximally fragmented Mott state rather than a fully fragmented one, $\Delta$ does not reach the ideal value of $0.11$, which would correspond to a perfectly 9-fold fragmented Mott phase.

\subsection{Commensurate filling factor $\nu = 1$, melting of Mott correlation in the fully fragmented Mott-insulator phase}

In this section, we consider the setup of strongly correlated Mott with commensurate filling, consisting of $N = 9$ bosons and $S = 9$ lattice sites. Keeping the previous case of commensurate filling, the interaction strength is increased to $g_0 = 0.5 E_r$ and the primary lattice depth is also increased to $V_p = 10 E_r$. This leads to fully fragmented Mott insulator in the primary lattice (Fig.~\ref{fig:density}(c)), where the many-body state is described as $\vert 1, 1, 1, 1, 1, 1, 1, 1, 1 \rangle$. The correlation dynamics is monitored following a sudden introduction of the secondary lattice. We find the system remains robust even for sufficiently strong disorder, and all the essential features of the primary lattice are preserved throughout the dynamics.

\begin{figure}[tbh]
    \centering
    \includegraphics[width=1.0\columnwidth]{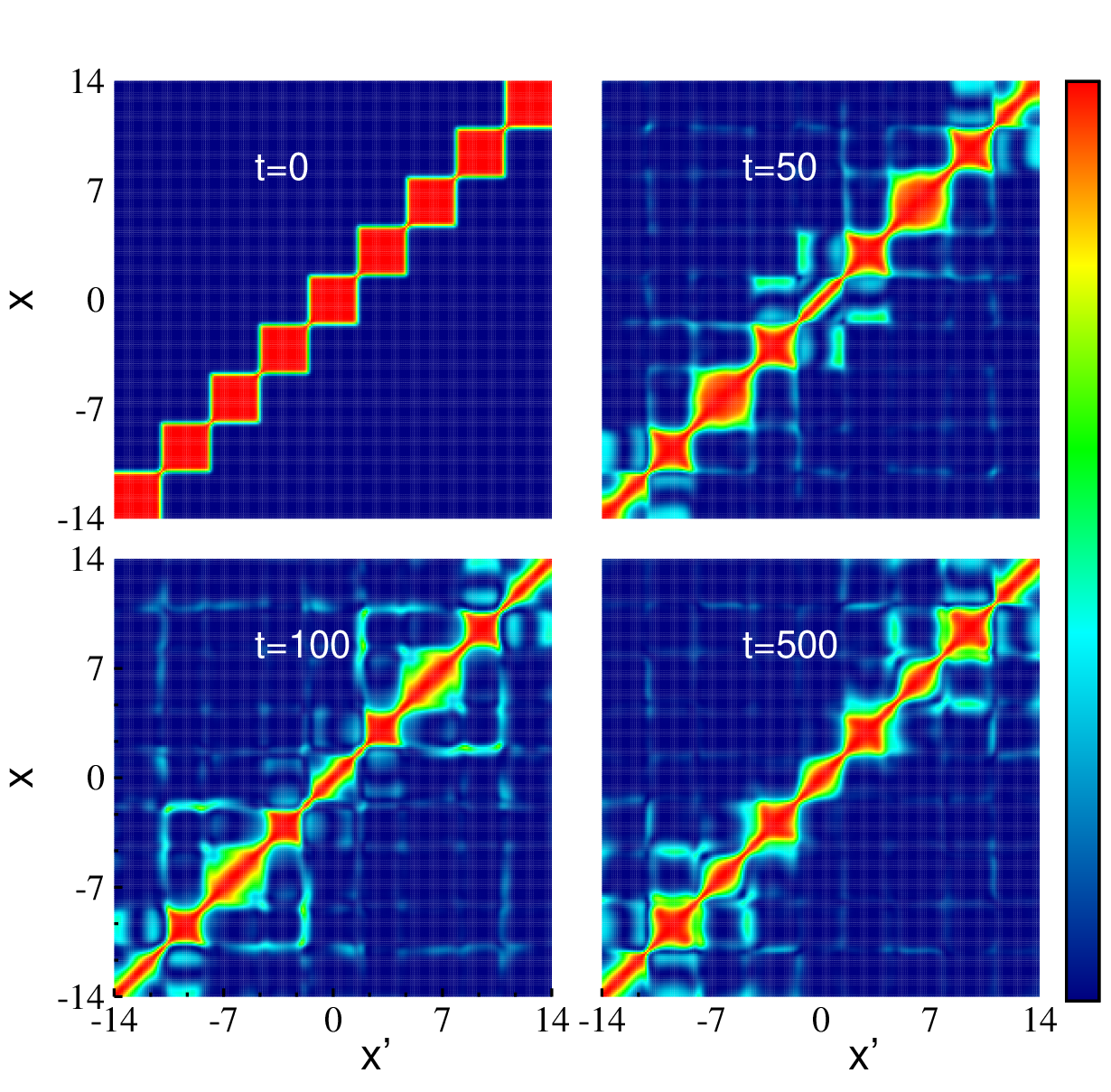}
    \caption{Dynamics of Glauber one-body correlation function $g^{(1)}(x, x')$ post quench for $N = 9$ bosons in $S = 9$ lattice sites. The computation is done with $M = 10$ orbitals. The other parameters, $V_p = 10 E_r$, $g_0 = 0.5 E_r$, $V_d = 3 E_r$. At $t = 0$, in the prequench state, the diagonal shows nine completely separated and highly coherent lobes with $|g^{(1)}|^2 \simeq 1$ which signifies fully localized Mott correlation. With time, disorder is able to distort the Mott correlation only, without leading to a new phase. See the text for details.}
    \label{fig:corr-1B-MI}
\end{figure}

In Fig.~\ref{fig:corr-1B-MI}, we present the dynamics of normalized one-body correlation function $g^{(1)}(x,x')$ following a strong disorder quench with strength $V_d = 3 E_r$. At time $t = 0$, the diagonal of $g^{(1)}(x,x')$ exhibits nine completely separated and highly coherent lobes with $|g^{(1)}|^2 \approx 1$, while the off-diagonal correlations ($x \ne x'$) vanish. This indicates a nine-fold fragmented many-body state. As time evolves, the secondary lattice begins to deplete the lattice coherence, resulting in the gradual melting of the Mott lobes. The initially fully coherent Mott phase in the primary lattice becomes partially incoherent, allowing for some inter-well tunneling. This partial delocalization of Mott coherence leads to phase distortion, although certain sites remain intact. Even in the final time of computation, $t=500$, we observe that the strong disorder competes with the initial Mott correlation. We do not observe any new dynamical phase, disorder only introduces distortions in phase coherence of Mott phase.

%

\begin{figure}[tbh]
    \centering
    \includegraphics[width=1.0\columnwidth]{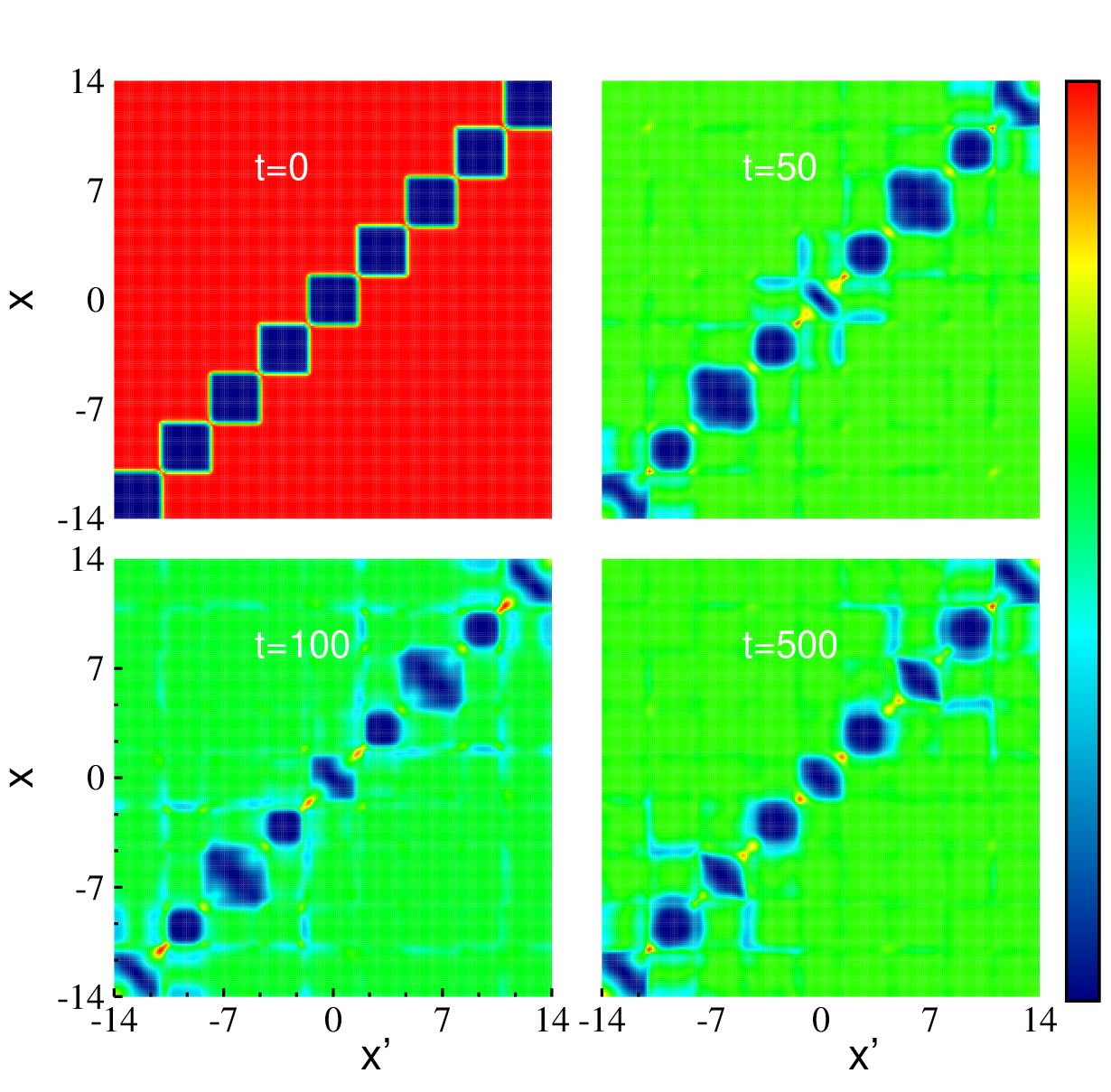}
    \caption{Dynamics of Glauber two-body correlation function $g^{(2)}(x, x')$ following a quench for the same parameters as in Fig~\ref{fig:corr-1B-MI}. At $t = 0$, $g^{(2)}$ exhibits nine dark lobes along the diagonal signifying the fully coherent fragmented Mott state. In the dynamics, disorder distorts the correlation hole and the off-diagonal correlation is slightly depleted. See the text for details.}
    \label{fig:corr-2B-MI}
\end{figure}

In Fig.~\ref{fig:corr-2B-MI}, we plot the two-body corelation function $g^{(2)}(x, x')$ for the same time point as for one-body correlation (Fig.~\ref{fig:corr-1B-MI}). At $t = 0$, the nine extinguished lobes (correlation hole) along the diagonal and complete off-diagonal correlation exhibits nine-fold fragmented Mott phase. In the dynamics, disorder competes with the Mott correlation in a very complex way. Compared to the previous case of commensurate filing, here the prequench state is strongly correlated. As a result, during the dynamical evolution, the secondary lattice simply induces partial distortions in the correlation holes and leads to a depletion of off-diagonal correlations. However, the overall stability of the Mott state is largely preserved.

\begin{figure}[tbh]
    \centering
    \includegraphics[width=1.0\columnwidth]{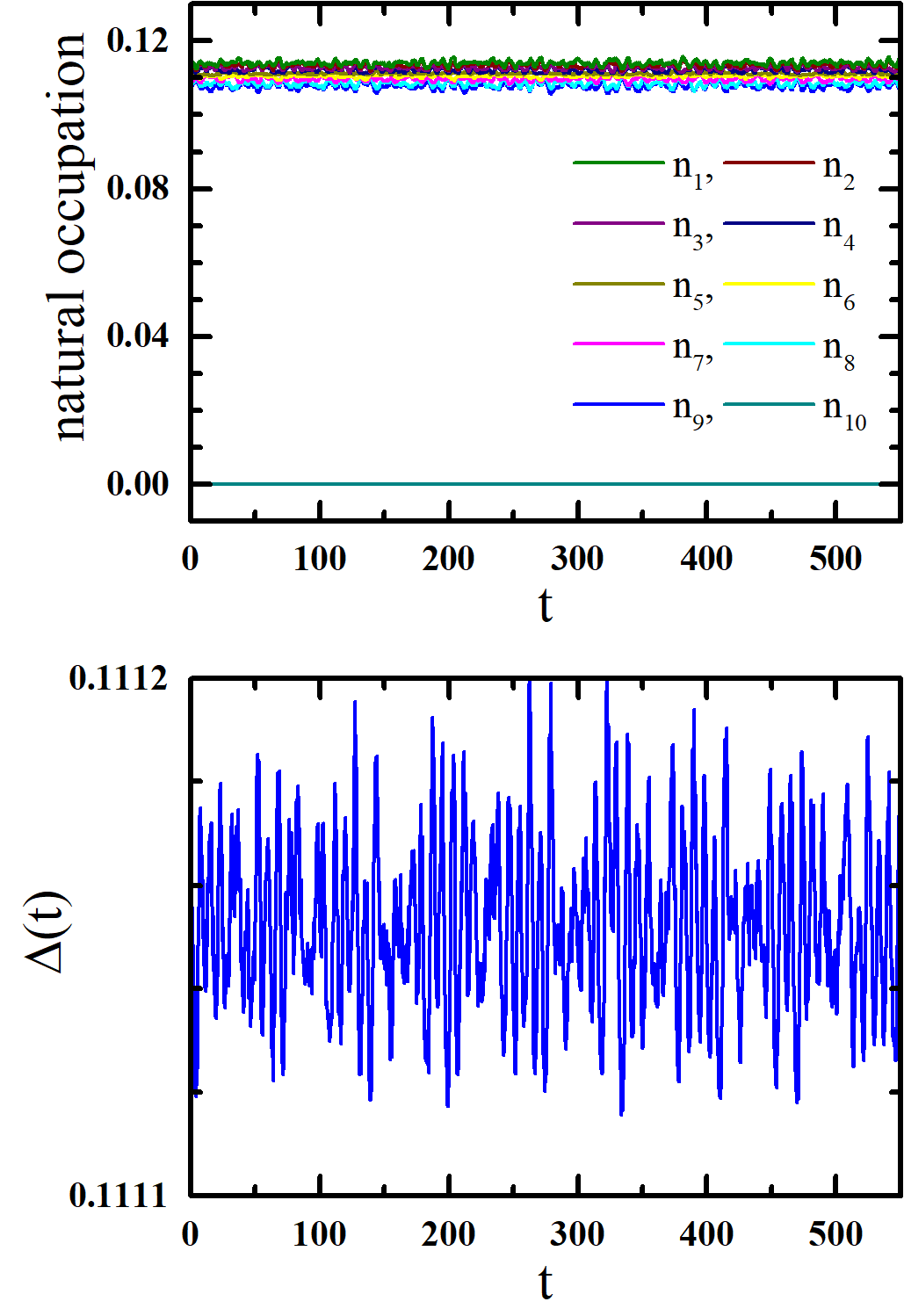}
    \caption{Top panel: Dynamics of orbital fragmentation when the fully fragmented Mott state following a quench for the same parameters as in Fig.~\ref{fig:corr-1B-MI} and Fig.~\ref{fig:corr-2B-MI}. Number of orbitals used in the computation is $M = 10$. The entire dynamics exhibits $S = 9$ fold fragmented Mott as each of $M = 9$ orbitals contribute equally ($n_1 \simeq n_2 \simeq n_3 \simeq n_4 \simeq n_5 \simeq n_6 \simeq n_7 \simeq n_8 \simeq n_9 = \frac{1}{9}$). Bottom panel: Dynamics of the order parameter, which fluctuates around the mean value of $0.11$, signifying the presence of a fully fragmented Mott phase throughout the entire dynamics. See the text for details.}
    \label{fig:fragmentation-MI}
\end{figure}

The dynamical fragmentation is presented in the top panel of Fig.~\ref{fig:fragmentation-MI}. The results clearly show that fully fragmented Mott state in the prequench configuration, characterized by an equal, 11.11\% occupation of each of $M = 9$ orbitals, remains fully fragmented in the entire dynamics. The last orbital has no contribution in the quench dynamics. Small fluctuations around the natural occupation value of $0.11$ for each of the $M = 9$ orbitals reflect minor distortions in the Mott correlations induced by the quench. In the bottom panel, we plot the dynamics of order parameter, which fluctuates about the expected value of $\Delta = \frac{1}{(S = 9)} \approx 0.11$, thereby confirming that the system retains its fully fragmented Mott character throughout the dynamics.

\subsection{Commensurate filling factor $\nu = 2$, resilience of fermionized Mott insulator phase}

\begin{figure}[tbh]
    \centering
    \includegraphics[width=1.0\columnwidth]{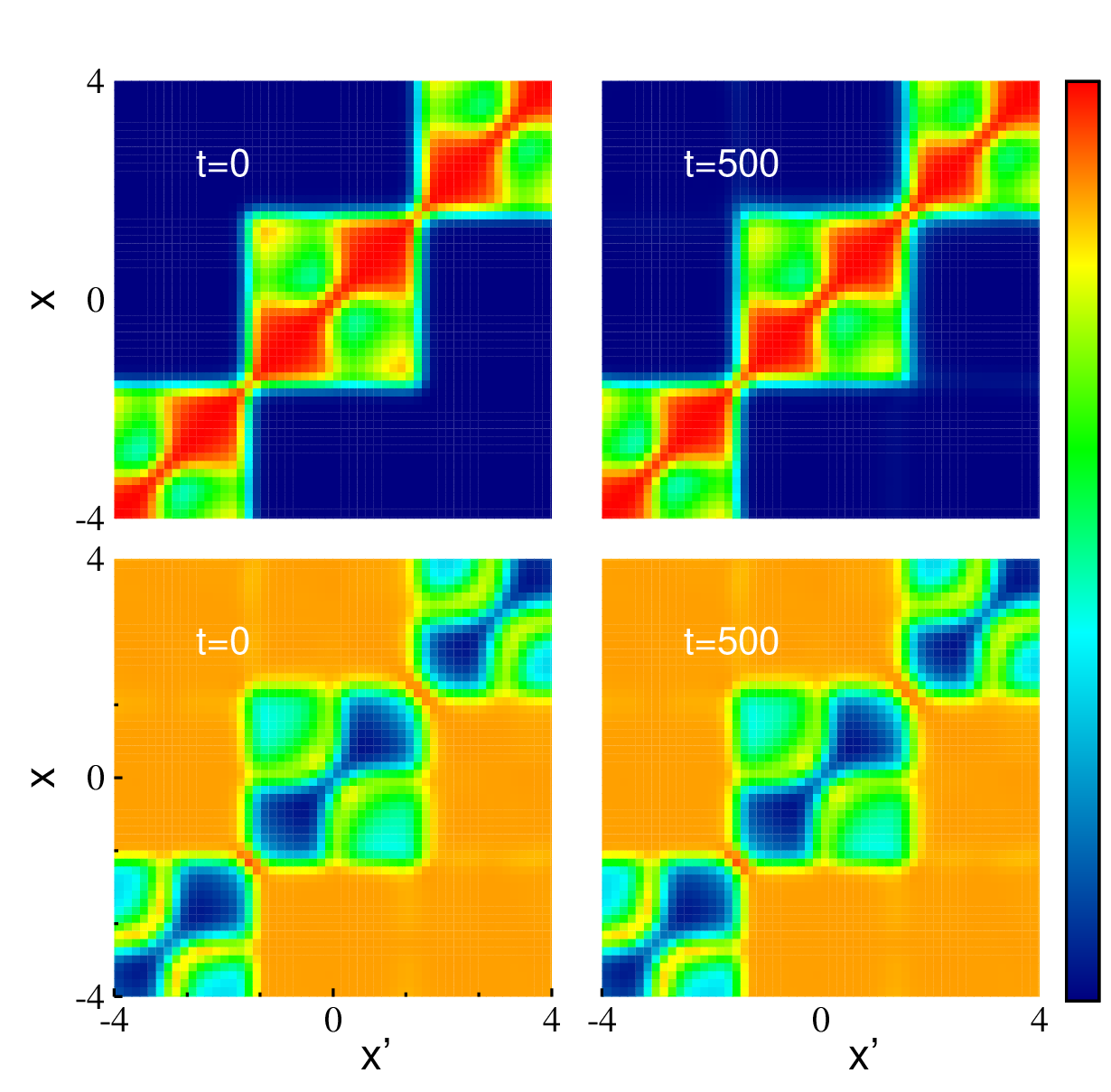}
    \caption{Top panel: Dynamics of Glauber one-body correlation function $g^{(1)}(x, x')$ post quench for $N = 6$ bosons in $S = 3$ lattice sites. The computation is done with $M = 12$ orbitals. The other parameters, $V_p = 15 E_r$, $g_0 = 10 E_r$, $V_d = 5 E_r$. At $t = 0$, in the prequench state, two bright lobes in each well exhibits spatially separated strongly interacting bosons. Bottom panel: Dynamics of Glauber two-body correlation function $g^{(2)}(x, x')$; two correlation holes in each well signifies strongly interacting dimer. The dimer correlation remains intact even in the strong disorder quench $(t = 500)$. See the text for details.}
    \label{fig:corr-1B-2B-FMI}
\end{figure}

Fig.~\ref{fig:corr-1B-2B-FMI} presents the dynamics of one- and two-body correlation functions when strongly interacting Mott $(g_0 = 10 E_r)$ in the deep lattice $(V_p = 15 E_r)$ with double filling (Fig.~\ref{fig:density}(d)) is suddenly quenched by a secondary lattice of strength, $V_d = 5 E_r$. In the strongly interacting limit, capturing the complete many-body correlation is challenging and requires significantly large number of orbitals in the computation. To ensure well-converged results, we reduce the number of lattice sites to $S = 3$. To observe dimer interactions in each site, we consider a strongly interacting system of $N = 6$ bosons. The computation is performed with $M = 12$ orbitals. The many-body state is configured as $\vert \left[1, 1\right], \left[1, 1\right], \left[1, 1\right] \rangle$. In terms of fragmentation, the pair of interacting particles occupy two orbitals in the same well. We present correlations only of the prequench state $(t = 0)$ and postquench state at the final point of the simulation $(t = 500)$. The top panel plots the one-body correlation $g^{(1)}(x, x')$. At $t = 0$, two distinct bright lobes in each well signifies the presence of two fermionized bosons. Two bosons in each site are now spatially separated dimer mimicking the onset of fermionization. Under a strong disorder quench, the dimer correlation remains unchanged. The bottom panel presents the two-body corelation function $g^{(2)}(x, x')$ for the same time point. At $t = 0$, two distinct correlation holes in each site again indicate the presence of two fermionized bosons, which remain robust in the strong disorder quench.

\section{Conclusion} \label{sec:conclusion}

In the present work, we compute dynamical response of strongly correlated many-body phases of periodic lattice upon sudden introduction of secondary lattice of incommensurate period. We investigate the out-of-equilibrium dynamics of four many-body phases ---non-fragmented and fragmented superfluids, fully fragmented Mott insulators, and fermionized Mott insulators in the quasi-periodic potential. The interplay between initial correlations and correlations induced by the secondary lattice governs the localization process, with MCTDHB capturing the distinctive role of dynamical fragmentation. Non-fragmented superfluids display collapse-revival localization, fragmented superfluids dynamically evolve towards Mott localization, and strongly correlated Mott phases remain largely robust, with only minor distortions. Fermionized Mott insulators preserve intra-dimer correlations despite off-diagonal perturbations. This work provides a comprehensive picture of many-body correlation dynamics, from uncorrelated to highly correlated phases, and introduces coherence measures as an independent signature of localization, offering an alternative experimental probe. Our findings serve as a guideline for exploring exotic localization phenomena arising from the interplay between short-range interactions and correlated disorder, and open perspectives for extending the approach to dipolar bosons, where long-range interactions may dynamically stabilize localized states in controlled disorder.

\section*{Acknowledgments}

This work was supported by the São Paulo Research Foundation under Grants No. 2013/07276-1, No. 2024/04637-8, and No.2025/00547-7, and by the National Council for Scientific and Technological Development under Grant No. 386392/2024-2. Texas A\&M University is
acknowledged.

\appendix

\section{System parameters}

\begin{table}[tbh]
    \centering
    \begin{tabular}{ || c | c || }
        \hline \hline
            Quantity & MCTDH-X units  \\
        \hline \hline
            unit of length & $\bar{L} = \frac{\lambda_p}{3} = 344$~nm \\
        \hline
            unit of energy & $\bar{E} = \frac{\hbar^2}{2 m \bar{L}^2} = E_r (\frac{3}{\pi})^2$ \\
        \hline
            potential depth & $V = 10.0 \bar{E} \approx \: 9.128 E_r$ \\
        \hline
            on-site repulsion & $\lambda = 0.5 \bar{E} \approx \: 0.456 E_r$ \\
        \hline \hline
    \end{tabular}
    \caption{Units used in MCTDH-X simulations. $E_r = \frac{\hbar^2 k_p^2}{2m}$ is the recoil energy.}
    \vspace{-5pt}
\end{table}

The quasi periodic lattice is the superposition of two optical lattices; a primary lattice of depth $V_p$ and wavelength $\lambda_p$ and a secondary lattice of depth $V_d$ and wavelength $\lambda_d$ parameterized as:
\begin{equation}
    V(x) = V_p\sin^2(k_px) + V_d \sin^2(k_dx),
\end{equation}
where $k_i$ is the wave vector. We choose $\lambda_p \simeq 1032$~nm and $\lambda_d \simeq 862$~nm, which are compatible with real experimental realizations in ultracold atomic gases. These give vectors $k_p \simeq 6.088 \times 10^6$~m$^{-1}$ and $k_d \simeq 7.289 \times 10^6$~m$^{-1}$. We impose hard-wall boundaries to restrict the optical lattice to the central wells.

\subsection{Lengths}
In MCTDH-X simulations, we choose to set the unit of length $\bar{L} \equiv \frac{\lambda_p}{3} = 344$~nm, which makes the minima of the primary lattice appear at integer values in dimensionless units, while the maxima are located at half integer values. $x = 0$ is the center of the lattice which can host an odd number of lattice sites $S$.

\subsection{Energies}
The unit of energy $\bar{E}$ is defined in terms of the recoil energy of the primary lattice, i.e. $E_r \equiv \frac{\hbar^2 k_p^2}{2m} \simeq 3.182 \times 10^{-26}$~J with $m\simeq 38.963$~u, the mass of $^{39}$K atoms. Thus we define the unit of energy as $\bar{E} \equiv \frac{\hbar^2}{(2m L^2)} = E_r(\frac{3}{\pi})^2 = 2.904 \times 10^{-26}$~J. In typical experiments with quasi periodic optical lattice, the depth of the primary lattice is varied in the around few tens of recoil energies and the depth of the secondary lattice is varied around few recoil energy. In our simulations, we probe similar regimes: $V_p = \in \left[5 E_r, \, 15 E_r\right]$ and $V_d \in \left[0.2 E_r, \, 5 E_r\right]$. The on-site interactions are kept fixed $g_0 = 0.001 E_r$ and $0.01 E_r$ for weakly interacting; $g_0 = 0.05 E_r$ for stronger interaction and $10.0 E_r$ for very strong interaction. These values are experimentally accessible in ultracold quantum simulators.

\subsection{Time}
The unit of time is defined from the unit of length as $\bar{t} \equiv \frac{2m \bar{L}^2}{\hbar}$ = $\frac{2m \lambda_p^2}{\hbar} = 0.1306 \times 10^{-4}~\text{s} = 13.07~\mu\text{s}$.

\bibliography{biblio}

\end{document}